\documentclass[amssymb,prb,twocolumn,floats,amsmath]{revtex4}

\usepackage[T1]{fontenc}
\usepackage{bm}
\usepackage{graphicx}
\usepackage{amssymb}
\usepackage{amsfonts}
\usepackage{amsmath}
\usepackage{epstopdf}
\usepackage{color}

\begin{document}

\title{Effect of magnetic field on the lasing threshold of a semimagnetic polariton condensate}

\newcommand*{\fuw}{\affiliation{Institute of Experimental Physics, Faculty of Physics, University of Warsaw, ul. Pasteura 5, PL-02-093 Warsaw, Poland}}

\author{J.-G.~Rousset}
\email{J-G.Rousset@fuw.edu.pl}
\fuw
\author{B.~Pi\k{e}tka}
\email{Barbara.Pietka@fuw.edu.pl}
\fuw
\author{M.~Kr\'ol}
\fuw
\author{R.~Mirek}
\fuw
\author{K.~Lekenta}
\fuw
\author{J.~Szczytko}
\fuw
\author{W.~Pacuski}
\fuw
\author{M.~Nawrocki}
\fuw


\begin{abstract}
We evidence magnetic field triggered polariton lasing in a microcavity containing semimagnetic quantum wells. This effect is associated with a decrease of the polariton lasing threshold power in magnetic field.  The observed magnetic field dependence of the threshold power systematically exhibits a minimum which only weakly depends on the zero-field photon-exciton detuning. These results are interpreted as a consequence of the polariton giant Zeeman splitting which in magnetic field: leads to a decrease of the number of accessible states in the lowest polariton branch by a factor of two, and substantially changes the photon-exciton detuning.
\end{abstract}

\maketitle


\section{Introduction}

\begin{figure}
    \includegraphics[width=0.8\linewidth]{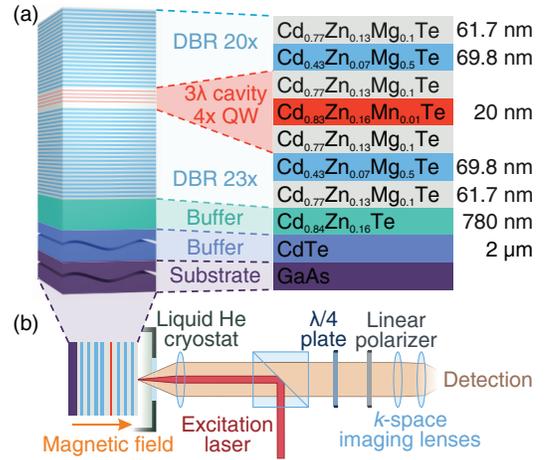}
  \caption{(a) Scheme of the sample designed for investigating semimagnetic polaritons. The layers are wedge-shaped with the average layer thicknesses given. The gradient of thickness is of the order of $2.5\%/$~cm. (b) Simplified experimental setup.}
  \label{setup}
\end{figure}

In the last years, there has been a growing interest for the magneto-optical and spin polarization properties of cavity polaritons and their Bose-Einstein condensates. Theoretical and experimental investigations evidenced, among others, the suppression of superfluidity in magnetic field  and the quenching of the Zeeman effect for polariton Bose-Einstein condensates. \cite{Rubo_PLA2006, Kulakovskii_PRB2012, Fisher_PRL2014} More recently, the condensation of polaritons triggered by magnetic field and the phase transition between polariton lasing and photon lasing in magnetic field were investigated and interpreted in the frame of the shrinkage of the exciton wave function in magnetic field, the diamagnetic shift and the influence of magnetic field on free carriers and exciton diffusion. \cite{Kochereshko_SciRep2016,Kochereshko_Semiconductors2016} Another approach to study the magneto-optical properties of photonic structures takes advantage of the enhanced magneto-optical properties of diluted magnetic semiconductors (DMS). It was shown that the magneto-optical Kerr effect can be amplified by placing a DMS layer on a distributed Bragg reflector. \cite{Cubian_PRB_2003,Sadowski_PRB_1997,Haddad_SSC_1999,Koba_JEWA2013, Koba_EPL2014, Pacuski_CDG2017} Magneto-optical effects of cavity polaritons can be enhanced by embedding quantum wells containing Mn ions in the cavity.\cite{UlmerTuffigo_JCG_1996,Ulmer_SLM97, Brunetti_PSSC_2005, Brunetti_PRB2006,Brunetti_PRB2006_spin_dynamics,Rousset_JCG2013, Rousset_JCG2014, Rousset_APL2015} The structure of the microcavity sample studied in this work is shown in Fig. \ref{setup} with the simplified experimental setup. The average layer thicknesses are given. In fact the microcavity is wedge-shaped with a gradient of thickness of the order of $2.5\%/$~cm. In such a structure, the exchange interaction between the $d$-shell electrons of Mn ions and the $s$-shell electrons and $p$-shell holes of the exciton results in an angle dependent giant Zeeman splitting of semimagnetic cavity polaritons. \cite{Mirek_PRB2017} In this work, we show that external magnetic field also affects the polariton properties in the non-linear regime of excitation. In particular we investigate the decrease of the polariton condensation threshold in magnetic field, which can lead to magnetic field induced lasing under constant excitation power.

\section{Experimental results}

Our sample consists of a 3$\lambda$-cavity embedding four 20~nm thick (Cd,Zn,Mn)Te quantum wells.\cite{Rousset_APL2015} The cavity is surrounded by distributed Bragg reflectors consisting of 22 pairs for the bottom mirror and 20 pairs for the top one.\cite{Rousset_JCG2013, Rousset_JCG2014} Polaritons are created non resonantly by a Al$_2$O$_3$:Ti pulsed laser which energy is tuned to the energy of the first reflectivity minimum on the high energy side of the microcavity stopband. All the presented results were obtained at a temperature of $T=1.5$~K.

\begin{figure*}
    \includegraphics[width=0.8\linewidth]{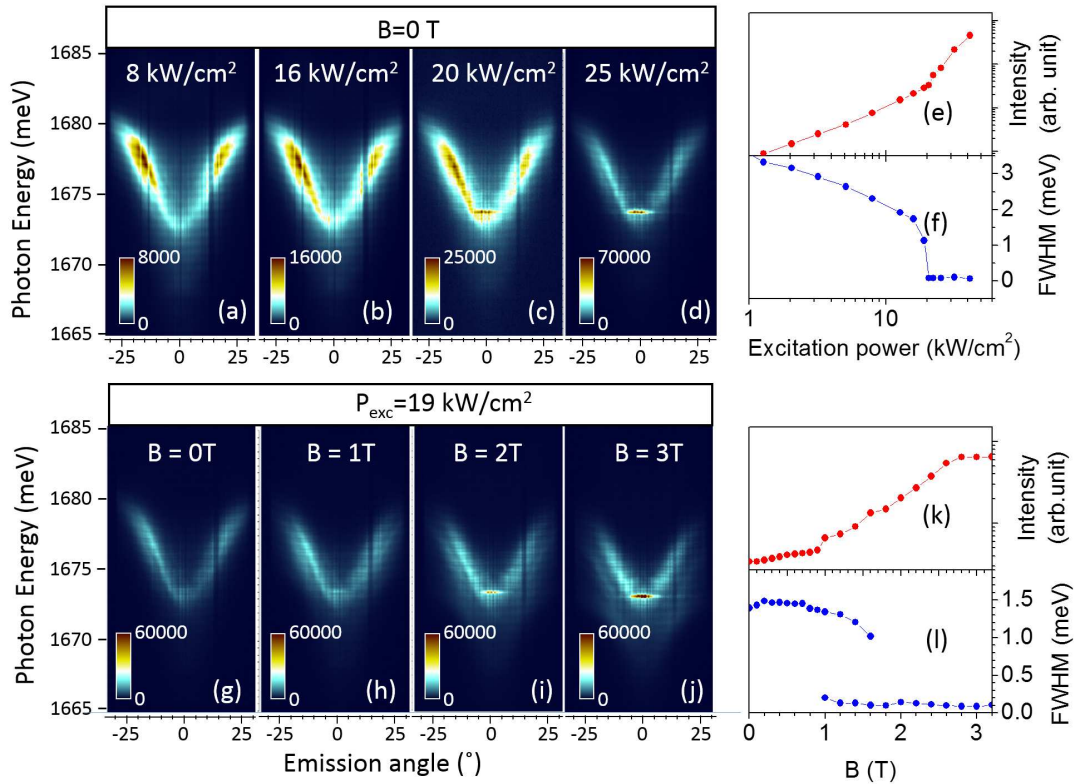}
  \caption{Comparison of two effects inducing polariton lasing: the increase of excitation power and the increase of magnetic field. (a)-(d) Angle resolved PL measurement of the polariton condensation in zero magnetic field when increasing the excitation power; the linear color scale is different for each panel. Panels (e) and (f) present the characteristic nonlinear intensity increase and linewidth narrowing at threshold. (g)-(j) Angle resolved PL measurements of the polariton condensation triggered by magnetic field under fixed excitation power; the same linear color scale is used for panels (g)-(j). Panels (k)  and (l) present the characteristic nonlinear intensity increase and linewidth narrowing at threshold. In panel (l), for $1$~T$\leq B \leq 1.6$~T, the emission at zero angle results from the contribution of two lines with comparable intensity: a narrow one from condensed polaritons and a broad one from uncondensed polaritons, which intensity relative to the narrow line becomes negligibly small above $B \approx 2$~T.}
  \label{dispersions}
\end{figure*}

To observe lasing triggered by magnetic field we apply the following procedure. In a first step, in zero magnetic field, the condensation threshold $P_{th}(B=0)$ is determined based on the dependence of the emission on the excitation power, as shown in Fig. \ref{dispersions} (a-f). Then the excitation power is set to $P_{exc}=0.75\ P_{th}(B=0)$ and the magnetic field is increased. For the situation presented in Fig. \ref{dispersions}, the photon-exciton detuning (defined as the energy difference between the uncoupled photon and exciton at zero in-plane momentum) in zero magnetic field is $\delta_0 = -8.5$~meV and the threshold power in zero magnetic field is $P_{th}(B=0)=25$~kW/cm$^2$. The magnetic field is increased, keeping the excitation power constant $P_{exc}=19$~kW/cm$^2$. When the magnetic field increases to $B\approx2$~T, we observe polariton condensation similarly to what is observed when increasing the excitation power. The dispersion curves at chosen magnetic fields for which the system is below or above threshold are presented in Fig.~\ref{dispersions} (g-l). We observe a typical narrowing of the emission in momentum and energy space when increasing the excitation power in zero magnetic field (Fig. \ref{dispersions} (e), (f)) or the magnetic field at a fixed excitation power (Fig.~\ref{dispersions}~(k),~(l)), which is characteristic of cavity polariton condensation and lasing.

\begin{figure}
    \includegraphics[width=0.8\linewidth]{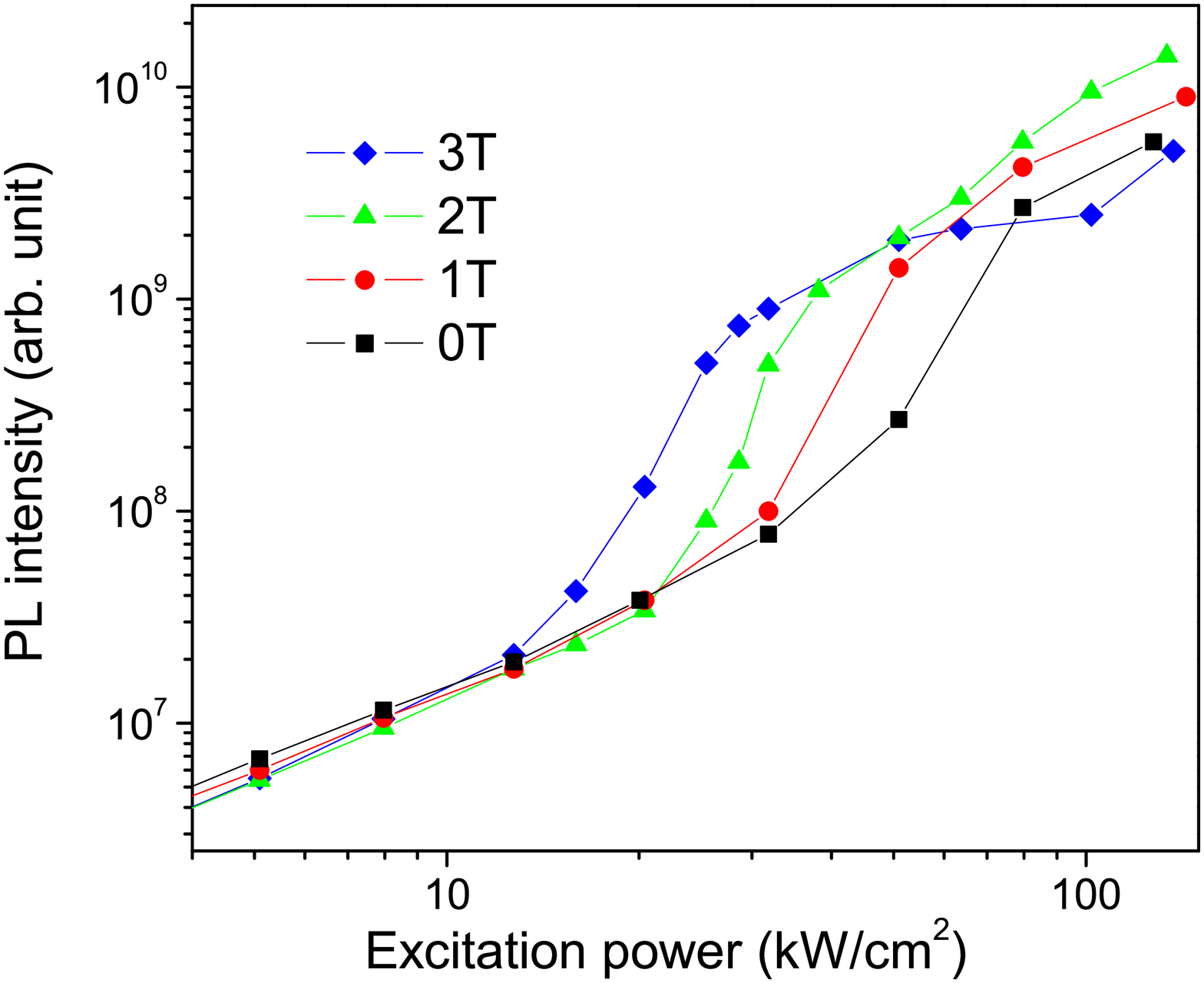}
  \caption{PL intensity \emph{vs} excitation power for various magnetic fields. When the magnetic field increases from $B=0$~T to $B=3$~T, the polariton lasing threshold power decreases.}
  \label{I-PvsB}
\end{figure}

Investigation of the PL intensity dependence on the excitation power, in various magnetic fields from $B=0$~T to $B=3$~T was performed. As shown in Fig. \ref{I-PvsB}, the non linear increase of PL intensity is observed and the condensation threshold power decreases with magnetic field.

The polariton condensation triggered by magnetic field shown in Fig. \ref{dispersions} and the associated decrease of the lasing threshold power led us to investigate the dependence of the polariton lasing threshold on  magnetic field and photon-exciton detuning. Series of measurements performed for four different values of the photon-exciton detuning starting from positive $\delta_0 = +3.0$~meV, close to zero $\delta_0 = -1.5$~meV, and significantly negative $\delta_0 = -2.4$~meV and $\delta_0 = -5.0$~meV are presented in Fig.\ref{Pth_vs_B}. For a given value of the detuning, PL intensity measurement vs excitation power was performed several times at various magnetic fields between $0$~T and $10$~T. This allowed us to determine the lasing power threshold as a function of magnetic field. For each detuning, the threshold power decreases in magnetic field compared to the $B=0$~T case.

\section{Discussion}

Previous zero-field studies show that the polariton lasing threshold power strongly depends on the shape of lower polariton dispersion curve which is determined by the photon-exciton detuning. \cite{kasprzak_PRL2008, Wertz_APL2009} Two types of related phenomena are considered: first - kinetic ones, such as the change of the polariton lifetime, relaxation processes and their efficiency, and second - thermodynamic ones, such as the change of the polariton effective mass, number of accessible states at the bottom of the lower polariton dispersion curve and the effective lattice temperature. \cite{kasprzak_PRL2008} Investigation of the condensation threshold dependence on the photon-exciton detuning show\cite{kasprzak_PRL2008, Wertz_APL2009} that when increasing the detuning from negative to positive values, the lower polariton branch forms a shallower trap in reciprocal space. In a shallower trap, polaritons encounter less scattering events to relax to the bottom of the trap and the threshold power decreases due to a more efficient relaxation of polaritons to the lowest energy state. For positive detuning, the lower polariton branch is predominantly excitonic and a further increase of the detuning increases the effective mass of the polaritons and the number of accessible states on the lower polariton branch close to $k_{||}=0$. This makes it harder for polaritons to reach an occupancy close to unity at the bottom of the lower polariton branch  which is the condition to trigger the stimulated scattering to the lowest energy state, and consequently increases the threshold power. A trade off between the kinetic limitation of the threshold at negative detuning and thermodynamic limitation at positive detuning, minimizes the threshold power close to zero detuning.\cite{kasprzak_PRL2008, Wertz_APL2009}

In microcavities with semimagnetic quantum wells presented in this work, the photon-exciton detuning, but also the density of states depend strongly on magnetic field.\cite{Mirek_PRB2017}  Consequently we consider here five mechanisms which may be involved in the magnetic field dependence of the polariton lasing threshold power:

\textbf{Effect of excitonic shift in magnetic field.} (i) The giant Zeeman splitting and related redshift of the low energy excitonic component leads to a more positive photon-exciton detuning and increases the efficiency of polaritons relaxation along the lower polariton dispersion curve, similarly to what is observed for nonmagnetic polaritons when increasing the detuning from negative to zero.\cite{kasprzak_PRL2008, Wertz_APL2009} This effect shall be important for negative photon-exciton detuning. (ii) The increase of the photon-exciton detuning induced by magnetic field  increases the effective mass and the number of accessible states for polaritons close to $k_{||}=0$,  making it harder for polaritons to reach occupancy close to unity and trigger the stimulated scattering to the lowest energy state. This effect shall be important for positive photon-exciton detuning.

\textbf{Effect of spin degeneracy and relaxation dynamics.} (iii) The magnetic field induced giant Zeeman splitting of polaritons lifts the spin degeneracy and consequently decreases by a factor of two the number of thermally accessible states\cite{Leggett_RevModPhys} in the lowest polariton branch compared to the $B=0$~T case,  when the $\sigma^+$ and $\sigma^-$ spin polarizations are degenerate. We propose that this reduced density of states makes it easier for polaritons to reach occupancy close to unity. (iv) The magnetic field accelerates spin-lattice relaxation of magnetic ions\cite{Strutz_PRL92,Dietl_PRL1995,Scherbakov_PRB2000,Goryca_PRL2009QW,Goryca_PRB2015} and excitons\cite{Smolenski_PRB2015}, therefore we propose that it could also accelerate the dynamics of polaritons towards the lowest energy state. (v) The shrinkage of the exciton wave function in magnetic field results in an increase of the oscillator strength and consequently a higher Rabi energy, which lowers the condensation threshold. Such a dependence on the Rabi energy of the condensation threshold has been studied \emph{e.g.} for wide gap semiconductor microcavities.\cite{Johne_APL2008, Jamadi_PRB2016}

\begin{figure}
    \includegraphics[width=1\linewidth]{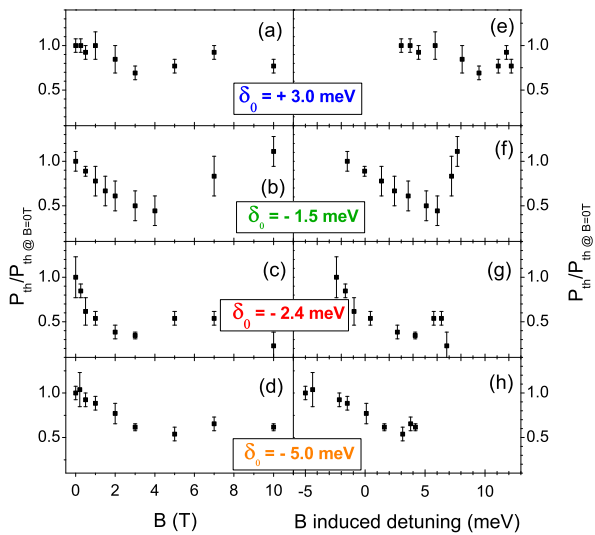}
  \caption{(a)-(d) Magnetic field dependence of the threshold power normalized to the value in zero magnetic field for various values of the photon-exciton detuning in zero magnetic field $\delta_0$. (e)-(f) The same data presented as a function of the photon-exciton detuning taking into account the excitonic Zeeman effect.}
  \label{Pth_vs_B}
\end{figure}

In order to discuss the importance of all these mechanisms, we analyze the dependence of the polariton condensation threshold power on magnetic field for various values of zero magnetic field detuning $\delta_0$ (Fig. \ref{Pth_vs_B}). The excitonic giant Zeeman effect leads to a symmetrical splitting which shifts the excitonic components to lower ($\sigma^+$ component) and higher ($\sigma^-$ component) energy, but owing to the low temperature at which the experiments are conducted, only the lower component is significantly occupied. Therefore the giant Zeeman effect leads to an increase of the photon-exciton detuning for the occupied branch of polaritons. This means that for negative $\delta_0$, the absolute value of the detuning decreases with magnetic field, whereas it increases with magnetic field for positive $\delta_0$.
From these considerations, if one assumes that the magnetic field induced changes of detuning (mechanism (i) and (ii)) are the most important factor governing the dependence of the threshold power on magnetic field, one would expect that for negative $\delta_0$ the lasing threshold should decrease with magnetic field, whereas it should increase with magnetic field for positive $\delta_0$. However, this is not what we observe in the data presented in Fig. \ref{Pth_vs_B}(a-d), where the lasing threshold power decreases in magnetic field, independently of the zero field detuning $\delta_0$. Moreover, plotting our data as a function of the magnetic field induced detuning (including the excitonic Zeeman effect), we observe that zero-detuning seems not to be the optimal condition for condensation in magnetic field (\ref{Pth_vs_B}e-h), which differs from what is found for nonmagnetic polaritons, for which the condensation threshold power is minimized close to zero detuning. This clearly means that other effects should be seriously taken in to account to explain the observed decrease of the lasing threshold power in magnetic field.

\begin{figure}[t]
    \includegraphics[width=0.8\linewidth]{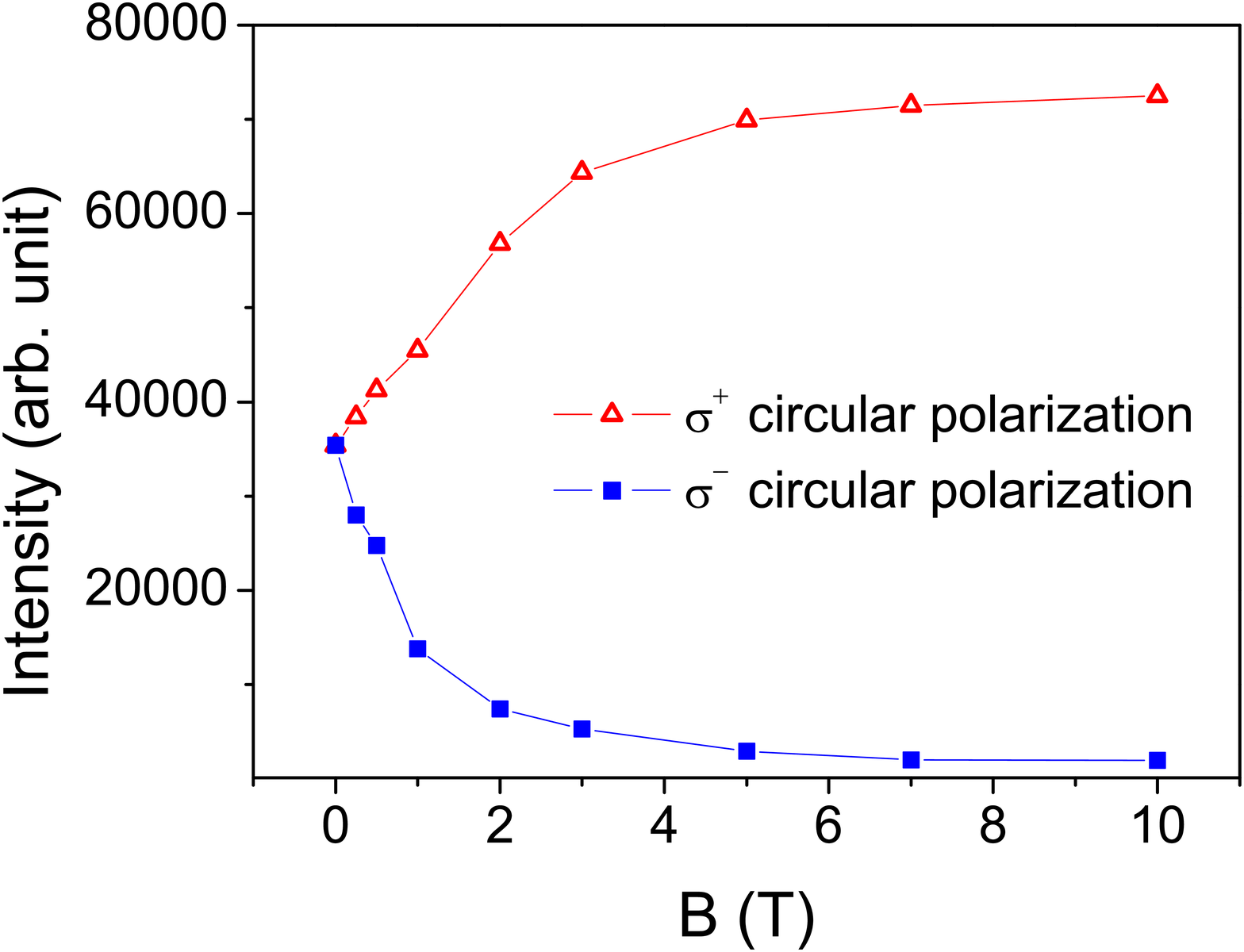}
  \caption{Magnetic field dependence of the PL intensity near $k_{||}=0$ collected in $\sigma^+$ and $\sigma^-$ circular polarization. The excitation power is above the lasing threshold in zero magnetic field.}
  \label{DOCP}
\end{figure}

We notice that the minimum threshold power is roughly half of the threshold power in zero magnetic field (Fig. \ref{Pth_vs_B}). This is a strong argument in favor of mechanism (iii) related to the lifting of spin degeneracy in magnetic field. Indeed, for $B \approx 4$~T, the spin polarization of the lower polariton near $k_{||}=0$ is close to saturation (as shown in Fig. \ref{DOCP}). Then, for $B \approx 4$~T the number of accessible states at the bottom of the lower polariton branch (in the vicinity of $k_{||}=0$) is roughly half the one in zero magnetic field. As a consequence, the number of polaritons needed for the onset of stimulated scattering to the lowest energy state, is also roughly decreased by a factor of two.

Comparing the four series of measurements shown in Fig. \ref{Pth_vs_B}, we observe that the minimum value of the threshold power $P_{th}^{min}$ (normalized with respect to the threshold in zero magnetic field) is lower for negative zero-field detuning $\delta_0$ than for positive detuning $\delta_0$.  This dependence can be understood in terms of an interplay between mechanisms (i), (ii), and (iii): when $\delta_0$ is negative, the magnetic field induced increase of detuning makes the polaritons relaxation kinetics more efficient which additionally decreases the threshold power (mechanism (i)). When $\delta_0$ is positive, the magnetic field induced increase of detuning increases the effective mass and density of states in the vicinity of $k_{||}=0$ which tends to increase the threshold power (mechanism (ii)), but the influence of this latter mechanism is weaker than effect of the reduced density of states (mechanism (iii)) resulting in a decrease of the threshold power. This interplay between mechanisms (i), (ii), and (iii) can also explain the increase of the threshold power for higher magnetic fields.

If the decrease of the lasing threshold is due to enhanced polariton relaxation in magnetic field, similarly to spin-lattice relaxation (mechanism (iv)), then it would be even more efficient for high magnetic fields,\cite{Goryca_PRB2015} which is not confirmed by the results presented in Fig. \ref{Pth_vs_B}. Therefore, such a mechanism cannot be of major importance in our experiment. The same argument applies for other effects involving the direct influence of magnetic field on the exciton motion.\cite{Kochereshko_SciRep2016}

Concerning the effect of the shrinkage of the exciton wave function in magnetic field leading to an increase of the coupling strength (mechanism (v)), or to a decrease of the exciton lifetime,\cite{Kochereshko_SciRep2016} owing to the small Bohr radius of excitons in CdTe QWs, such an effect is expected~\cite{Kavokin_MC2007} to be non negligible for magnetic fields higher than $25$~T which is far above the range of magnetic fields applied in these investigations.\\

\section{Conclusions}

The effect of polariton lasing induced by external magnetic field for microcavity containing semimagnetic quantum wells is observed. We show that this effect is due to the decrease of the polariton lasing threshold power in magnetic field. From the analysis of the threshold power dependence on magnetic field for various values of the photon-exciton detuning we can infer that the mechanism leading to a decrease by roughly a factor two of the threshold power in magnetic field is the reduction by a factor of two of the number of accessible states in the vicinity of the lowest energy state of the $\sigma^+$ polarized lower polariton dispersion curve. Since the efficiency of the threshold power decrease by magnetic field is higher for negative detuning than for positive detuning, lasing threshold is governed by an interplay between magnetic field induced changes of the density of states and photon-exciton detuning. The ability to control the operation of a polariton laser by a magnetic field opens new possibilities for combining fields of spintronics and polaritonics.

\section*{Acknowledgments}

The authors acknowledge Jan Suffczy\'{n}ski and Micha{\l} Matuszewski for helpful discussions. The work was supported by the National Science Centre, Poland, under Projects No. 2014/13/N/ST3/03763, No. 2015/16/T/ST3/00506, No. 2015/18/E/ST3/00559, No. 2015/18/E/ST3/00558, and by the Polish Ministry of Science and Higher Education as a research project Diamentowy Grant No. 0010/DIA/2016/45 in years 2016-2020 and No. 0109/DIA/2015/44 in years 2015-2019. Research was carried out with the use of CePT, CeZaMat and NLTK infrastructures financed by the European Union - the European Regional Development Fund within the Operational Programme "Innovative economy" for 2007-2013.

\end{document}